\documentclass{article}

\usepackage{amsthm}

\makeatletter
\renewcommand\thesubsection{\thesection.\@arabic\c@subsection}
\makeatother

\newcommand {\beq}{\begin{equation}}
\newcommand {\eeq}{\end{equation}}
\newcommand {\beqa}{\begin{eqnarray}}
\newcommand {\eeqa}{\end{eqnarray}}         
\newcommand {\beqs}{\begin{eqnarray*}}
\newcommand {\eeqs}{\end{eqnarray*}}
\newcommand {\bds}{\begin{displaymath}}
\newcommand {\eds}{\end{displaymath}}
\newcommand {\n}{\nonumber\\}

\newcommand {\bebb}{\begin{thebibliography}}
\newcommand {\eebb}{\end{thebibliography}}      
\newcommand {\bbit}{\bibitem}

\newcommand{\e}{{\epsilon}}

\def\l{\lambda}

\def\pd{\prod}

\def\ra{\rangle}

\def\dg{\dagger}

\def\journal#1&#2(#3){\unskip, \sl #1\ \bf #2 \rm(19#3) }
\def\andjournal#1&#2(#3){\sl #1~\bf #2 \rm (19#3) }

\usepackage{cite}

\begin{document}
\vskip 1cm
\begin{flushright}
\end{flushright}

\vspace{1cm}

\begin{center}
{\Large\bf Quasi-Exactly Solvable Models Derived from 
the Quasi-Gaudin Algebra}

\vskip.1in

{\large Yuan-Harng Lee, Jon Links,  and Yao-Zhong Zhang }
\vskip.1in

{\em School of Mathematics and Physics,
The University of Queensland, Brisbane, Qld 4072, Australia}

\end{center}

\begin{abstract}
The quasi-Gaudin algebra was introduced to construct 
integrable systems which are only quasi-exactly solvable. 
Using a suitable representation of the quasi-Gaudin algebra, we obtain a class of 
bosonic models which exhibit this curious property. These models have the 
notable feature that they do not preserve $U(1)$ symmetry, which is typically associated to a non-conservation of particle number. 
An exact solution for the 
eigenvalues within the quasi-exactly solvable sector is obtained via the algebraic Bethe 
ansatz formalism.   
\\ \n 
PACS Numbers: 02.30.Ik, 03.65.Fd, 05.30.Jp.
\end{abstract}

\section{Introduction}

In \cite{Ushveridze97, Ushveridze98} Ushveridze proposed a method for
studying  quasi-exactly solvable (QES) systems \cite{Turbiner88,qes,Ushveridze94}
from the prespective of integrable systems and the Quantum Inverse Scattering Method
(QISM) \cite{Korepin}. The approach, which is called the {\it partial} algebraic Bethe 
ansatz (ABA), relies on deforming the Yang-Baxter algebra in such a way that it retains
most of the features required for the QISM but leads to generating functions of 
integrable systems which are only QES. This deformation of the Yang-Baxter algebras led to
new classes of hitherto unknown algebras. A limiting case is the (rational) quasi-Gaudin algebra which 
will be the focus of this study.  \\ \n 
Exactly solvable models have found many successes in various branches of physics and mathematics. 
Over recent years they have continued to find new applications in diverse fields such as  
Bose-Einstein condensates and degenerate Fermi gases, quantum optics, superconductivity, and nuclear pairing 
among other things e.g. \cite{Ortiz05,Foerster07,Amico07,draayer09,Amico10,bfgk10,dilsz10,llz11}. 
There has also been significant interest in QES models, with new applications of these being found in problems relating to matrix product states \cite{mps}, and in
dissipative systems \cite{jarvis}. However by comparison the partial ABA approach seems to have received 
little attention and remains essentially undeveloped. 
An appealing property of the partial ABA is that it provides us with a constructive
algebraic approach for obtaining QES models which have multiple degrees of freedom.
\\\n 
One particular aspect of the ABA which bears some relevance to our present 
exposition is the study of quantum integrable models which do not preserve $U(1)$ 
symmetry. Such models are interesting for a number of reasons. In the context of 
spin-boson Hamiltonians of the Tavis-Cummings form, these models correspond to physical systems without the 
rotating wave approximation.
Diagonalisation of such models 
is a somewhat complicated affair within the ABA method due to the 
lack of reference states, often requiring the use of functional Bethe ansatz or 
Sklyanin's separation of variable technique  \cite{Amico07,Amico10}. 
Non $U(1)$ preserving 
models are also relevant to the study of open quantum systems whereby the $U(1)$ symmetry 
is broken due to coupling to an environment. An example of this is found in the spin-boson Hamiltonian of Leggett et al 
\cite{leggett} which has found applications ranging from quantum-state engineering \cite{mss} to biomolecular systems \cite{ross}. 
\\\n 
In the present paper, we will study QES bosonic models descending 
from suitable realisations of the quasi-Gaudin algebra. It will be shown that such 
models correspond to an extension of the $su(1,1)$ Dicke Hamiltonian \cite{Tik08} by 
the addition of U(1) symmetry-breaking terms. The Hamiltonian can be written as 
\beqa H= H_0 + H_1 \eeqa 
with $H_0$ and $H_1$ refering to the Dicke Hamiltonian and the U(1) symmetry-breaking 
component respectively. Explicitly, they have the form   
\beqa 
H_0 &=& wN_b+\sum_{i=1}^m 2\e_iS_i^z + g \left(\sum_{i=1}^m b S_i^{+} + b^{\dg}S_i^- \right),\n 
H_1 &=&  g\left((b + b^{\dg})\left(n+f^z-\sum_{i=1}^m S_i^z\right) -b^{\dg}b^2 - (b^{\dg })^2b \right). 
\eeqa 
Here $N_{b},b,b^{\dg} $ are standard bosonic operators, $f^z$ is a representation 
dependent parameter, $w_0,\e_i,g$ are free parameters, $n$ is an integer and  
$S_i^{z,\pm}$ are either single-mode or double-mode representations of $su(1,1)$ 
generators (refer to equations (\ref{brep1}) and (\ref{brep2}) below).  The Hamiltonian $H_1$ may be 
interpreted as a coupling of the $su(1,1)$ Dicke model to an external system. 
\n \\
Our paper is structured as follows. In Section 2 we will briefly review the partial 
ABA method of obtaining quasi-exact solutions for models associated to the 
quasi-Gaudin algebra. In Section 3 we will use a suitable representation of the 
quasi-Gaudin algebra to obtain the integrable bosonic model (1.1). We then derive the 
Partial ABA solution of the Hamiltonian and discuss aspects relating to the quasi-exact solvability. Finally in Section 4 we  summarise our results and discuss possible 
future lines of work.    

\section{Quasi-Gaudin Algebra and Bethe Ansatz Solution}
 
Let us first introduce the rational (rank 1) Gaudin algebra and the associated 
abstract, integrable models before defining its quasi counterpart. The rational 
Gaudin model is a parameter-dependent infinite-dimensional Lie algebra satisfying the 
following commutation relations:  
\begin{eqnarray*}
S^z(\lambda)S^z(\mu)
-S^z(\mu)S^z(\lambda)&=&0, \nonumber\\[0.4cm]
S^\pm(\lambda)S^\pm(\mu)
-S^\pm(\mu)S^\pm(\lambda)&=&0, \nonumber\\[0.3cm]
S^z(\lambda)S^\pm(\mu)
-S^\pm(\mu)S^z(\lambda)&=&
\pm\frac{S^\pm(\lambda)-S^\pm(\mu)}{\mu-\lambda},\nonumber\\
S^-(\l)S^+(\mu)-S^+(\mu)S^-(\l)
&=&
2\frac{S^z(\lambda)-S^z(\mu)}{\mu-\lambda},
\end{eqnarray*}
whereby $\l$ and $\mu$ are complex spectral parameters. From these relations, it 
can be shown that
\beqa 
H(\l) = S^z(\l)S^z(\l) -\frac{1}{2}S^+(\l)S^-(\l)-\frac{1}{2}S^-(\l)S^+(\l)
\eeqa  
satisfies the following commutation relations   
\beqa 
\left[ H(\l), H(\mu) \right] = 0 
\eeqa 
and therefore acts as a generator of commuting operators in an abstract integrable system. Assuming the existence of a suitable reference state, the 
spectrum of $H(\l)$ can be obtained via the standard ABA \cite{Ushveridze98}.
 
Analogous to the Gaudin algebra is the so-called quasi-Gaudin algebra. 
It is defined by the following parameter dependent set of 
relations \cite{Ushveridze97,Ushveridze98}   
\beqa
S_n^z(\lambda)S_n^z(\mu)
-S_n^z(\mu)S_n^z(\lambda)&=&0, \nonumber\\[0.4cm]
S_{n\pm 1}^\pm(\lambda)S_n^\pm(\mu)
-S_{n\pm 1}^\pm(\mu)S_n^\pm(\lambda)&=&0, \nonumber\\[0.3cm]
S_{n\pm1 }^z(\lambda)S_n^\pm(\mu)
-S_n^\pm(\mu)S_{n}^z(\lambda)&=&
\pm\frac{S_n^\pm(\lambda)-S_n^\pm(\mu)}{\mu-\lambda},\nonumber\\
S_{n+1}^-(\lambda)S_n^+(\mu)
-S_{n-1}^+(\mu)S_n^-(\lambda)&=&
2\frac{S_n^z(\lambda)-S_n^z(\mu)}{\mu-\lambda} \label{qGaudRel}
\eeqa
whereby $n$ is an integer and $\l$, $\mu$ are complex parameters. While 
(\ref{qGaudRel}) appears to be similar to the Gaudin algebra,  we stress 
that there are important qualitative difference between the two. Importantly,   
note that (\ref{qGaudRel}) do not define commutation relations and are 
therefore not Lie algebraic relations. Despite looking somewhat arbitary, the quasi-Gaudin 
algebra can be understood as a grading deformation on the original Gaudin algebra. 
We refer the reader to \cite{Ushveridze98} for a more 
detailed discussion.   

Similar to the Gaudin algebra, there exists a generating function of commuting operators  
for the quasi-Gaudin algebra. It has the form
\beqa 
H_n(\l) = S_n^z(\l)S_n^z(\l)  -\frac{1}{2}S^-_{n+1}(\l)S^{+}_n(\l)-\frac{1}{2}S^+_{n-1}(\l)S^-_n(\l)
\eeqa 
and can be shown to form a commutative family with respect to the spectral parameters, 
i.e.  
\beqa 
\left[ H_n(\l), H_n(\mu)\right]=0. \label{qgaudinint}
\eeqa 
Note that the commutation relation (\ref{qgaudinint}) does not extend to the general 
case where $H_n(\l)$ and $H_m(\mu)$ have different integer values of $n$ and $m$. This is due 
to the lack of a defining relations between elements of the algebra with arbitrary 
integer indexes. The ABA solution for the generating function $H_n(\l)$ of the quasi-Gaudin algebra has 
been obtained in \cite{Ushveridze97,Ushveridze98}.  As wtih the standard Gaudin algebra, the ABA 
diagonalisation of $H_n(\l)$ works if the representation of (\ref{qGaudRel}) 
supports a reference state $|0\ra $, viz. 
\beqa 
S_0^z(\l)|0\ra = f(\l)|0\ra  ~~,~~ S_0^-(\l)|0\ra = 0
\eeqa 
The Bethe vector is given by 
\beqa 
\psi(\mu_1, \cdots \mu_n) = S_{n-1}^+(\mu_n)S_{n-2}^+(\mu_{n-1})\cdots S_{0}^+(\mu_1)|0 \ra .\label{qBethev}
\eeqa 
By successively applying the following relation
\beqa 
H_n(\l) S^+_{n-1}(\mu_n) = S^+_{n-1}(\mu_n)H_{n-1}(\l) + 2\frac{S^+_{n-1}(\mu_n)S^z_{n-1}(\l)- S^+_{n-1}(\l)S^z_{n-1}(\mu_n)}{\l-\mu_n}
\eeqa 
we can shift the operator $H_n(\l)$ towards the right of the product of $S^+_i(\mu_{i+1})$ operators on the right-hand side of (\ref{qBethev}), so that we 
finally have $H_n(\l)$ acting on the reference state.  After having completed 
this procedure, we perform the same operation for the various $S^z_i(\l),\,S^z_i(\mu_{i+1})$ that were 
generated as a byproduct of shifting the $H_n(\l)$ through the product of the $S^+_{i}(\mu_{i+1})$. 
The final form is given by 
\beqa 
H(\l)\psi(\mu_1, \cdots \mu_n) = A(\l)\psi(\mu_1, \cdots \mu_n) + 2\sum_i B(\mu_i)\psi(\mu_1, \cdots, \mu_{i-1}, \l, \mu_{i+1}, \cdots ,\mu_n) 
\eeqa 
whereby 
\beqa 
A(\l) &=& f(\l)^2+ f'(\l) + 2\sum_{i=1}^n \frac{f(\l)}{\l-\mu_i} +2\sum_{i=1}^n\frac{1}{\l-\mu_i}\sum_{j \ne i}\frac{1}{\mu_i -\mu_j},\n 
B(\mu_i) &=& f(\mu_i) + \sum_{j \ne i} \frac{1}{\mu_i- \mu_j}. 
\eeqa 
By requiring that the unwanted terms vanish we obtain the following Bethe ansatz equations:
\beqa 
\sum_{k=1, k \ne i}^n\frac{1}{\mu_i- \mu_k} + f(\mu_i)=0, \qquad i=1,2,...,n
\eeqa 
with the eigenvalue for $H_n(\l)$ given by 
\beqa 
E_n(\l)= f^2(\l)+ f'(\l)+ 2 \sum_{i=1}^n \frac{f(\l)- f(\mu_i)}{\l-\mu_i}.
\eeqa  
As a proof of existence, an explict representation for (\ref{qGaudRel}) is provided in 
\cite{Ushveridze97,Ushveridze98}:  
\beqa 
S^-_n(\l) &=& S^-(\l) + \frac{f^z-S^z + n}{\l-c},\n 
S^0_n(\l) &=& S^0(\l) +\frac{f^z-S^z + n + d}{\l-c},\n 
S^+_n (\l) &=& S^+(\l) +\frac{f^z-S^z + n+2d}{\l-c}. \label{qGaudinRep}
\eeqa  
with  $c$ and $d$ as free parameters, $S^{\pm,z}(\l)$ are generators of the  
Gaudin algebra, and $S^z$ and $f^z$ are defined as 
\beqa 
S^z= \lim_{\l\rightarrow \infty}\l S^z(\l), \label{qGaudinRes}~~ S^z|0\ra = f^z|0\ra . 
\eeqa 
In terms of this realisation, the generating function $H_n(\l)$  takes the form
\beqa 
H_n(\lambda)&=&S^z(\lambda)S^z(\lambda)-\frac{1}{2}
S^-(\lambda)S^+(\lambda)-\frac{1}{2}S^+(\lambda)S^-(\lambda)\nonumber\\
&&\quad+\frac{2S^z(\lambda)(n+d+f^z-S^z)-S^-(\lambda)(n+2d+f^z-S^z)
-S^+(\lambda)(n+f^z-S^z)}{\lambda - c}\nonumber\\
&&\quad -\frac{1}{4(\lambda-c)^2}.  \label{gGaudinGF}
\eeqa  
It can be seen that the condition of hermiticity for (\ref{gGaudinGF}) is satisfied when 
$d={1}/2$ and the representation for the Gaudin algebra is 
unitary, i.e. satisfying the condition
\beqa 
S^+(\l)^{\dg} = S^-(\l),~~~~ S^z(\l)^{\dg}= S^z(\l). 
\eeqa

\section{Bosonic Representations of the Quasi-Gaudin Algebra}

The quasi-Gaudin algebra of the form (\ref{qGaudinRep}) admits  mixed representations, 
consisting of $su(1,1)$ algebras and the Heisenberg algebra, with the following form:
\beqa 
S^-_n(\l) &=& \frac{2b}{g}+\sum_{i=1}^m\frac{S_{i}^-}{\l-\epsilon_j} +\frac{f^z-N_b-\sum_i S_i^z+n}{\l-c}, \n 
S^z_n(\l) &=& \frac{w-2\l}{g^2}+\sum_{i=1}^m\frac{S_{i}^z}{\l-\epsilon_j}  +\frac{f^z-N_b-\sum_i S_i^z+n+\frac{1}{2}}{\l-c},\n 
S^+_n(\l) &=& \frac{2b^{\dg}}{g}+\sum_{i=1}^m\frac{S_{i}^+}{\l-\epsilon_j}  +\frac{f^z-N_b-\sum_i S_i^z+n+1}{\l-c}. \n 
\label{su11}
\eeqa 
The $S_i^{\pm,z}$ and $\{N_b, b,b^{\dg}\}$ are respectively the  
$su(1,1)$ and Heisenberg algebras, which obey the commutation relations
\beqa 
\left[ S^z_i, S^{\pm}_j \right] = \pm S^{\pm}_i\delta_{ij} ~,~ \left[S^-_i,S^+_j \right] = 2S_i^z\delta_{ij} \n 
\left[N_b,b^{\dg} \right] =b^{\dg} ~,~\left[N_b,b \right]= -b ~,~ \left[b,b^{\dg} \right] = 1
\eeqa 
and $S^z$ and $f^z$ are defined as  
\beqa 
S^z = \sum_{i=1}^m S_i^z + N_b, ~~~~ S^z|0\ra = f^z|0\ra .
\eeqa 
We note here that our definition for $S^z$ differs from that of (\ref{qGaudinRes}) as  
the prior definition is divergent for this particular realisation.  
 
The $su(1,1)$ algebras has two bosonic operator realisations. The first  is given by 
the single-mode representation, whereby 
\beqa 
S^z_i = \frac{a_i^{\dg}a_i}2+ \frac{1}{4} =\frac{N_{a_i}}2 + \frac{1}{4},~~~~ S^+_i= \frac{(a_i^{\dg })^2}{2} ~~,~~ S^-_i= \frac{a_i^2}2. \label{brep1}
\eeqa 
The second one is given by the two-mode representation, 
\beqa 
S^z_i = \frac{1}{2}\left( a_i^{\dg}a_i + c_i^{\dg}c_i \right) + \frac{1}2 =\frac{\left({N_{a_i}} + {N_{c_i}} \right)}{2}  +\frac{1}{2},~~~~ S^+_i= {a_i^{\dg}c_i^{\dg}}, ~~~~ S^-_i= {a_ic_i}. \label{brep2}
\eeqa 
There are multiple reference states for both  bosonic realisations. For the single-mode 
realisation, there are finitely many of them. We can express them as 
\beqa 
|0, \{l\}\ra = \pd_{i=1}^m (a_i^{\dg})^{l_i}|0\ra ~~,~~ l_i = 0~ \textrm{or} ~1
\eeqa  
where $\{ l\}$ is a shorthand notation for the set $\{l_1, \cdots l_m\}$ and 
\beqa 
S^z|0, \{l\}\ra = f^z|0, \{l\}\ra =\left(\sum_{i=1}^m \frac{l_i}{2} +\frac{1}{4} \right)|0, \{l\}\ra .
\eeqa 
For the two-mode realisation, there are infinitely many reference states. Without loss 
of generality we can write them as
\beqa 
|0, \{l\} \ra  = \pd_{i=1}^m (a_i^{\dg})^{ l_i}|0\ra , ~~~~ l_i =0,1,2, \cdots 
\eeqa 
with 
\beqa 
S^z|0, \{l\}\ra = f^z|0, \{l\}\ra =\left(\sum_{i=1}^m \frac{l_i}{2} +\frac{1}{2} \right)|0, \{l\}\ra .
\eeqa
It can be seen that each reference state corresponds to a distinct eigenfunction of the 
Casimir operators for the $su(1,1)$ generators $S_i^{\pm,z}$. As the $su(1,1)$ Casimir 
operators acts as central elements with respect to (\ref{brep1}) and (\ref{brep2}), we can use 
Schur's lemma to deduce that  each reference state gives rise to a distinct 
irreducible representation.

\section{Quasi-Exactly Solvable Hamiltonians } 
We now consider the generating function $H_n(\l)$ of the quasi-Gaudin algebra 
obtained from the representation  (\ref{su11}). Assuming $\e_i \ne \e_j$, it can be 
seen that 
\beqa 
H_n(\l)&=& -\frac{4}{g^2}\left(n+f^z+\frac{1}2\right)+\frac{1}{g^4}(w-2\l)^2 - \frac{2}{g^2}\left(
\frac{H_c}{\l-c}+\sum_{j=1}^m \frac{H_j}{\l-\e_j}  \right)  \nonumber\\
&&\quad + \sum_{i=1}^m\frac{K_i}{(\l-\e_i)^2} - \frac{1}{4(\l-c)^2} \label{qgaudinggf}
\eeqa 
with
\beqa 
H_j &=& ( 2\e_i-w)S_j^z + g\left( b^{\dg}S_j^- +bS_j^-\right)+ \sum^m_{i\ne j} 
\frac{1}{\e_j-\e_i}\left( 2S_i^zS_j^z - S_i^+S_j^- - S_i^-S_j^+\right) \n 
&& + \frac{g^2\left(S_j^z\left( n+\frac{1}2 +f^0-S^0\right)-\frac{1}{2} 
S_j^-(n+1+f^z-S^z)-\frac{1}{2}S_j^+(n+f^z-S^z)\right)}{\e_j-c}, \n 
H_c &=& g^2\sum^m_{i=1}\frac{S_i^z\left( n+ \frac{1}2 +f^z-\sum^m_{i=1} 
S_i^z-N_b\right)-\frac{1}{2}S_i^+(n+f^z-S^z) -\frac{1}{2}S_i^-(n+1+f^z-S^z)}{(c-\e_i)} 
\n &&+(2c-w)\left(n+\frac{1}{2}+f^0-\sum^m_{i=1}S_i^z-N_b \right)+ 
gb^{\dg}\left(n-\sum^m_{i=1}S_i^z-N_b \right) + gb\left(n+1-\sum^m_{i=1}S_i^z-N_b \right),\n     
K_i &=& S_i^zS_i^{z}-\frac{1}{2}\left( S_i^-S_i^+ + S_i^+S_i^-\right)  .
\eeqa
From (\ref{qgaudinggf}) and the commutation relation (\ref{qgaudinint}),  it follows 
that $H_{i,c}, K_{i,c}$ form a set of mutually commuting operators. By considering the 
following linear combination  $H= \Upsilon+ \sum_{i}H_i + H_c$ and setting the coefficient $c=0$, 
we obtained the desired bosonic hamiltonian. 
For the single-mode representations, we have 
\beqa 
H &=& wN_b+\sum_{i=1}^m \e_iN_{a_i}  +g\sum_{i=1}^m \left(b (a_i^{\dg})^2 + 
b^{\dg}a_i^2 \right)  
\n 
&& 
+ g \left((n+f^z)(b^{\dg}+b) - (b + b^{\dg})\sum_{i=1}^m \frac{N_{a_i}}{2} 
-b^{\dg}b^2 - (b^{\dg})^2b \right) \label{Hb1}
\eeqa
where 
$$\Upsilon =   w\left(n+\frac{1}{2} + f^z\right)-\sum_{i=1}^m \frac{\e_i}2.$$
For the two-mode representations, we obtain  
\beqa 
H &=& wN_b+\sum_{i=1}^m \e_i \left(N_{a_i}+N_{c_i} \right)+g\sum_{i=1}^m \left(b 
a_i^{\dg}c_i^{\dg} + b^{\dg}a_ic_i \right)  \n && + g \left((n+f^z)(b^{\dg}+b) - (b + b^{\dg})\sum_{i=1}^m 
\frac{N_{a_i}}{2} -b^{\dg}b^2 - (b^{\dg})^2b \right)
\eeqa
with 
$$\Upsilon =   w\left(n+\frac{1}{2} + f^z\right)-\sum_{i=1}^m {\e_i}.$$
We note that for the case when $m=1$, the models correspond to quasi-exactly solvable 
extensions for atom-molecule BEC models contained \cite{links03}. 

The eigenvalues for the Hamiltonians can be extracted from the Bethe ansatz solution of 
(\ref{qgaudinggf}):  
\beqa 
E_n(\l) &=& f^2(\l)+f'(\l)+2\sum_{i=1}^n\frac{f(\l)-f(\mu_i)}{\l-\mu_i}.
\eeqa 
This is done by evaluating the residues of the poles ${\e_i}$ and $c$. Doing so 
yields
\beqa 
E &=& \Upsilon-w\left( \sum_{i=1}^m s_i^z+ \frac{1}2\right)+\sum_{i=1}^m 2\e_is_i^z +  
\frac{g^2}2\left(\sum_{j=1}^m\sum_{i=1}^n\frac{s_j^z}{\mu_i-\e_j} +  \sum_{i=1}^n 
\frac{1}{2(\mu_i-c)} \right)     
\eeqa 
whereby $s_i^z = (2{l_i} + {1})/4$ for the single-mode representations  and  
$s_i^z = ({l_i} + {1})/2$ for the two-mode representations.

We now examine the quasi-exactly solvable nature of the Hamiltonians in more detail. 
For the sake of clarity, we shall only consider the Hamiltonian with the single-mode bosonic 
representation (\ref{Hb1}), as results for the two-mode representation will follow analogously. 
It is straightfoward to see that (\ref{Hb1}) acts on  an infinite-dimensional Hilbert 
space $V$ span by the following basis states
\beqa 
V \equiv \textrm{span}\{ (b^{\dg})^{ l_0}(a_1^{\dg})^{ l_1}\cdots (a_m^{\dg})^{ l_m}|0\ra \} \equiv  
\textrm{span} \{|l_0,\cdots, l_m\ra \}, ~~~~ l_i \in Z^+ .
\eeqa
In order to identify the invariant subspace which characterises the quasi-exact 
solvability of the Hamiltonian, let us write the Hamiltonian as  
\beqa 
H_g= H_0+H_-+H_+
\eeqa
whereby we have introduced a grading structure on the Hamiltonian through setting
\beqa 
H_0 &=& wN_b+\sum_{i=1}^m \e_iN_{a_i} + g \left(\sum_{i=1}^m b (a_i^{\dg})^{2} + 
b^{\dg}a_i^2 + (n+f^z)(b^{\dg}+b) \right), \n 
H_+ &=& (n+f^z)b^{\dg} -  b^{\dg}\sum_{i=1}^m \frac{N_{a_i}}{2}  - (b^{\dg})^2 b, \n 
H_- &=& (n+f^z)b - b\sum_{i=1}^m \frac{N_{a_i}}{2} -b^{\dg}b^2. \label{Hgraded}
\eeqa
The assigned grading of  ${\pm,0}$ is determined by the commutation relations of 
$H_{\pm,0}$ with the $U(1)$ charge $ S^z=N_b + \sum_{i=1}^m (2{N_{a_i}} + 
{1})/{4}  $:
\beqa 
\left[ S^z, H^0 \right] = 0 ~~,~~ \left[ S^z, H^+\right] = H^+ ~~,~~ \left[S^z, H^-
 \right]= -H^- .\label{U1Rel}
\eeqa 
In light of these relations, we may decompose $V$ into a direct sum of eigenspace 
$V_{i,p}$ of the $U(1)$ charge $ S^z$ and the Casimir operators of the $su(1,1)$ 
algebra $K_i =S_i^{z}(S_i^z-1)-S^+_iS^-_i $, i.e.  
\beqa 
V= \bigoplus_{i,\{p\}}V_{i,\{p\}}.
\eeqa 
Explicitly, the subspace $V_{i,\{p\}}$ can be written  as 
\beqa 
V_{i,\{p\}} \equiv \textrm{span}\{ (b^{\dg})^{ l_0}(a_1^{\dg})^{ 2l_1 +p_1}\cdots (a_m^{\dg})^{ 2l_m +p_m}|0\ra\}, ~~~~ \sum_{j=0}^m l_j  =i, ~~~~ p_i = 0 \textrm{ or } 1 .
\eeqa 
It can also be verified that  
\beqa 
S^z V_{i,\{p\}} = \left(i+\sum_{j=1}^m\left( \frac{p_i}2 + \frac{1}4 \right) \right)V_{i,\{p\}}, ~~~~ K_iV_{i,\{p\}} = \left( \frac{p_i}2 + \frac{1}4 \right) \left( \frac{p_i}2 - \frac{3}4 \right) V_{i,\{p\}}.
\eeqa 
From the commutation relations (\ref{U1Rel}), we therefore have
\beqa 
H_+ V_{i,\{p\}} \subseteq V_{i+1, \{p\}}, ~~~~ H_0V_{i, \{p\}} \subseteq V_{i, \{p\}}, ~~~~ H_- V_{i,\{p\}} \subseteq V_{i-1,\{p\}}.
\eeqa 
The QES property of the Hamiltonian arises from the fact that for given integer value of $n$ and 
$f^z=\sum_i\left( {l_i}+{1}\right)/4 $, we have 
$$
H_+V_{n, \{l\}} = \{0\} .
$$
As a result,  the Hamiltonian leaves  the following subspace invariant:
\beqa 
V_{\textrm{{\footnotesize QES}}} \equiv \bigoplus_{i=0}^n V_{i, \{l\}}.
\eeqa 
We can indeed verify that the Bethe vectors lie within this invariant subspace, by 
expanding the eigenvectors (\ref{qBethev}) explicitly.  It would be interesting to 
examine the possibility of obtaining exact solutions outside of this sector.

\section{Conclusion}
 
We've investigated a class of QES, integrable multi-mode bosonic 
models using the quasi-Gaudin algebra. We see that such models are obtained via a mixed 
representation consisting of commuting copies of $su(1,1)$, and the Heisenberg algebra. 
Integrable Hamiltonians were extracted from the generating 
function of commuting operators. A notable feature was that the QES 
Hamiltonians we obtain do not preserve $U(1)$ symmetry. 
We identified the QES sector of the Hamiltonian as the direct sum of the eigensubspaces 
of the $U(1)$ charge with eigenvalues no greater than $n$. 
\\ \n 
The ABA method leads to partial solutions of the Hamiltonians we've considered. Given the integrability of the Hamiltonian, in the sense that $H_n(\l)$ acts as a generator of conserved operators, it would  be interesting to explore the possibility of obtaining the entire spectrum via some 
other techniques. The dominating experience is that integrability and exact solvabilty go hand-in-hand. It is not apparent for these Hamiltonians whether the full spectrum is potentially accessible. 
\\ \n
Finally we note that due to the constraint arising 
from imposing hermiticity on the generating function $H_n(\l)$,  the quasi-Gaudin formalism is at present 
limited to cases based on underlying unitary representations of $su(1,1)$, or the Heisenberg algebra. 
It would be of interest to obtain representations of the
quasi-Gaudin algebra based on non-unitary (in particular finite-dimensional) representations of $su(1,1)$, which are also able to accomodate hermitian Hamiltonians. 
\\ \n 
{\bf  Aknowledgements} \\ \n 
This work was supported by the Australian Research Council through Discovery Projects DP11013434 and DP110101414. 

\bebb{99}
\bbit{Ushveridze97}
A. Ushveridze, Mod. Phys. Lett. {\bf A13} (1998) 281-292 
\bbit{Ushveridze98}
A. Ushveridze, Ann. Phys. {\bf 266} (1998) 81-134 
\bbit{Turbiner88}
A. Turbiner, Commun. Maths. Phys. {\bf 118} (1988) 467-474
\bbit{qes} 
V.V. Ulyanov and O.B. Zaslavskii, Phys. Rep. {\bf 216} (1992) 179-251 
\bbit{Ushveridze94}
A.G. Ushveridze, { \it Quasi-Exactly Solvable Models in Quantum Mechanics},
Institute of Physics Publishing, London, 1994.
\bbit{Korepin}
V.E. Korepin, N.M. Bogoliubov, and A.G. Izergin, 
{\it Quantum Inverse Scattering Method and Correlation Functions}, 
Cambridge University Press, 1993.
\bbit{Ortiz05}
G. Ortiz, R. Somma, J. Dukelsky, and S. Rombouts, Nucl. Phys. B {\bf 707} (2005) 421-457
\bbit{Foerster07}
A. Foerster and E. Ragoucy, Nucl. Phys. B {\bf 777} (2007) 373-403
\bbit{Amico07}
L. Amico, H. Frahm, A. Osterloh, and G.A.P. Ribeiro, Nucl. Phys. B {\bf 787} (2007) 283-300 
\bbit{draayer09}
F. Pan, M.-X. Xie, X. Guan, L.-R. Dai, and J.P. Draayer,
Phys. Rev. C {\bf 80} (2009) 044306
\bbit{Amico10}
L. Amico, H. Frahm, A. Osterloh, and T. Wirth, Nucl. Phys. B {\bf 839} (2010) 604-626
\bbit{bfgk10} M.T. Batchelor, A. Foerster, X.-W. Guan, and C.C.N. Kuhn, 
J. Stat. Mech.: Theor. Exp. (2010) P12014
\bbit{dilsz10}
C. Dunning, M. Iba\~nez, J. Links, G. Sierra, and S.-Y. Zhao, 
J. Stat. Mech.: Theor. Exp. (2010) P08025
\bbit{llz11}
Y.-H. Lee, J. Links, and Y.-Z. Zhang,
Nonlinearity {\bf 24} (2011) 1975-1986
 
\bbit{mps} 
M. Sanz, M.M. Wolf, D. P\'erez-García, and J.I. Cirac,
Phys. Rev. A {\bf 79} (2009) 042308
\bbit{jarvis}
S.H. Jacobsen and P.D. Jarvis, 
J. Phys. A: Math. Theor. {\bf 43} (2010) 255305


\bbit{leggett}
A.J. Leggett, S. Chakravaty, A.T. Dorsey, M.P.A. Fisher, A. Garg, and W. Zwerger,
Rev. Mod. Phys. {\bf 59} (1987) 1-85

\bbit{mss}
Y. Makhlin, G. Sch\"on, and A. Shnirman, 
Rev. Mod. Phys. {\bf 73} (2001) 357-400

\bbit{ross}
J. Gilmore and R.H. McKenzie,
J. Phys.: Condens. Matter {\bf 17} (2005) 1735–1746
\bbit{Tik08}
I. Tikhonenkov, E. Pazy, Y. B. Band, and A. Vardi, Phys. Rev. A {\bf 77} (2008) 063624

\bbit{links03}
J. Links, H.-Q. Zhou, R.H. McKenzie, and M.D. Gould, J. Phys. A: Math. Gen. {\bf 36} (2003) R63-R104

\eebb 
\end{document}